# Instant Image Denoising Plugin for ImageJ using Convolutional Neural Networks

**Varun Mannam**∗**, Yide Zhang, Yinhao Zhu, Scott Howard**

*Department of Electrical Engineering, University of Notre Dame, South Bend, IN 46556, USA*
∗*vmannam@nd.edu*

**Abstract:** We present a new convolutional neural network (CNN) based ImageJ plugin for fluorescence microscopy image denoising with an average improvement of 7.5 dB in peak signal-to-noise ratio (PSNR) and denoising instantly within 80 msec. © 2020 The Author(s)

**OCIS codes:** 100.4996, 170.2520.

## 1. Introduction

Fluorescence microscopy has enabled a dramatic development in modern biology [1]. The imaging speed of fluorescence microscopy is fundamentally limited by the signal to noise ratio (SNR). There are two general approaches to directly improve imaging SNR. First, the signal level can be increased by increasing the excitation optical power. However, the amount of power that a biological sample can receive is limited, and high power could result in fluorophore saturation. Second, noise can be reduced either by increasing pixel dwell time, exposure time or the number of images for averaging. However, increased exposure may cause photodamage to the sample and is not suitable for imaging real-time dynamics in living animals. In order to increase imaging speed without increasing excitation power, denoising algorithms can be used to overcome the fundamental physical limit to SNR. Although several denoising algorithms exist, they suffer from limited performance and slow computational speed. In this work, we present a novel Poisson-Gaussian denoising algorithm that has a better peak signal-to-noise ratio (PSNR) compared to the existing approaches. The new approach is based on deep neural network (with the U-Net architecture) using noisy fluorescence microscopy images as input [2] and therefore requires significantly less computation time compared to alternative techniques. We then extend this approach by developing an ImageJ plugin that can denoise almost any microscopy image. We show that the denoised images using our approach have an average PSNR improvement of 7.5 dB and can be computed "instantly" (i.e., in real-time) during imaging.

## 2. Methods

In this work, we train a deep neural network with 12,000 fluorescence microscopy (confocal, two-photon and widefield) images with different noise levels (raw and averages of 2, 4, 8, and 16 images) of bovine pulmonary artery endothelial (BPAE) cells, zebrafish embryos, and mouse tissues. To be specific, we retrained the Noise2Noise deep learning model [2] (without batch-norm) in the TensorFlow/Keras. During training, another noisy image in the same field of view (FOV) is used as the target image. This alleviates the need of the ground-truth (actual target) image.

Fig. 1 illustrates the average PSNR of the test-mix dataset [2] using our image denoising approach. There is an improvement of 7.5dB in the average PSNR of instant-denoised images compared to the average PSNR of noisy (raw) images in the test-mix data. Also, we observe that our ImageJ denoising performance in PSNR is better than conventional denoising methods (PURE-LET, VST+BM3D, and VST+NLM), and this comparison is tabulated in [2] [Table 2].

Using this trained model, we then create an Image Denoising plugin to denoise 2D (gray and color noisy images) and 3D (gray and color noisy images) fluorescence microscopy images in ImageJ and the source code is provided for this work [1].

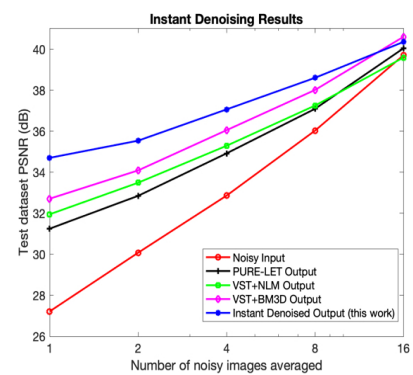

Fig. 1. Test dataset average PSNR

## 3. Results and Conclusions

We perform image denoising using our ImageJ plugin with the test-dataset given in [2]. Fig. 2 illustrates a sample mouse brain (2D grayscale) image captured using a two-photon excitation (TPE) microscope and a BPAE cell (2D color) image captured with a confocal microscope for noisy input, denoised output, and ground truth (target). In this work, the target image is generated by taking the average of 50 noisy images in the same FOV. We observe that our plugin reduces the Poisson-Gaussian noise by providing a better PSNR and the image also looks similar to the target image.

---
[1] https://github.com/varunmannam/Image_denoising/tree/master/Image_Denoising_Plugin.

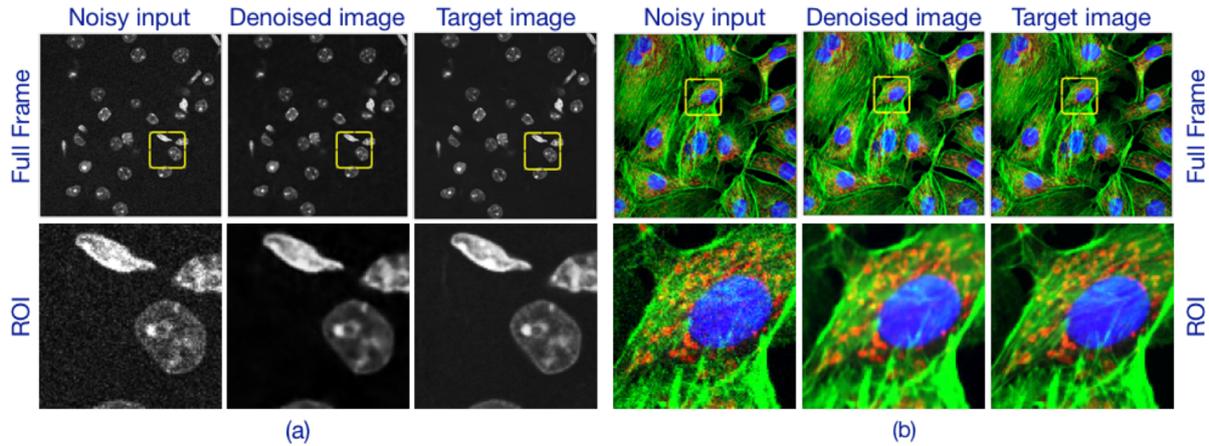

Fig. 2. Image denoising results using our ImageJ plugin: (a) Mouse brain captured with a two-photon microscope (pixel dwell time is 12 $\mu$s and the pixel width is 200 nm), (b) BPAE cell (prepared slide#1: F36924) captured with a confocal microscope (pixel dwell time is 12 $\mu$s and the pixel width is 100 nm); the top row indicates the full-frame (of size 512 $\times$ 512) of noisy input, denoised output, target and the bottom row indicates the region of interest (ROI: marked in the yellow square of size 100 $\times$ 100) from the respective top row images.

The PSNR of the denoised output from our approach is compared with the PSNR of conventional approaches and tabulated in Table 1. We considered two conventional approaches: ROF denoising [3] that performs 2D gray-scale images denoising using the nearest neighbors and Non-Local-Means denoise [4] (NLM denoise) method that performs 2D color images denoising using the non-local means algorithm. Our ImageJ denoising plugin provides an improvement of 9.95 dB and 5.23 dB in the PSNR in 2D gray and 2D color images respectively. In addition, the PSNR of our plugin based denoised image is comparable to taking the average of 8 noisy images (in the same FOV) and thereby reducing the computation time by 8 folds. The image denoising time in our ImageJ plugin using central processing unit (CPU) and graphics processing unit (GPU) are 960 ms and **80 ms** respectively. Therefore, image denoising speed is faster by 12 orders of magnitude when using GPU compared to using CPU. Overall, the ImageJ plugin can be used to denoise fluorescence images that are obtained with low laser power and a fast acquisition rate.

| Sample | Microscopy | PSNR | | |
|---|---|---|---|---|
| | | Noisy Input | **Instant Denoise Output** | Comparison Plugins |
| Mouse brain (2D gray) | Two-Photon | 24.9514 | **34.9085** | 33.4978 (ROF denoising [3]) |
| BPAE cell (2D color) | Confocal | 30.5034 | **35.7359** | 33.2834 (NLM denoising [4]) |

Table 1. Comparison of Instant denoising tool using PSNR to the existing ImageJ denoising plugins

In summary, we present a new denoising technique that allows imaging 8 times faster than the fundamental speed limit of fluorescence microscopy based on deep learning. The system was trained using 12,000 experimentally obtained images using the new Noise2Noise approach to develop a neural network that effectively removes Poisson-Gaussian noise present in widefield, confocal, and two-photon fluorescence microscopy images.